\documentclass[twocolumn,prl,showpacs]{revtex4}

\usepackage{graphicx}
\usepackage{bm}

\begin{document}

\title{Role of Oxygen Electrons in the Metal-Insulator Transition in the
Magnetoresistive Oxide La$_{2-2x}$Sr$_{1+2x}$Mn$_2$O$_7$ Probed by
Compton Scattering}

\author{
B. Barbiellini$^1$, A. Koizumi$^2$, P. E. Mijnarends$^{1,3}$, W.
Al-Sawai$^1$, Hsin Lin$^1$, T. Nagao$^2$, K. Hirota$^4$, M.
Itou$^5$, Y. Sakurai$^5$, and A. Bansil$^1$}

\affiliation{
$^1$Physics Department, Northeastern University, Boston MA 02115 \\
$^2$Graduate School of Material Science, University of Hyogo
 3-2-1 Kouto, Kamigori-cho, Ako-gun, Hyogo 678-1297, Japan\\
 $^3$Department of Radiation, Radionuclides \& Reactors,
 Faculty of Applied Sciences,
 Delft University of Technology, Delft, The Netherlands \\
 $^4$ Department of Earth and Space Science, Graduate School of Science,
 Osaka University 1-1 Machikaneyama, Toyonaka 560-0043, Osaka, Japan \\
 $^5$Japan Synchrotron Radiation Research Institute (JASRI),
 SPring-8, 1-1-1 Kouto, Sayo, Sayo, Hyogo 679-5198, Japan\\}

 \date{\today}

 \pacs{71.30.+h,75.47.Lx,78.70.Ck}


 \begin{abstract}
 We have studied the [100]-[110] anisotropy of the Compton profile in the
 bilayer manganite. Quantitative agreement is found between theory and
 experiment with respect to the anisotropy in the two metallic phases
 (i.e. the low temperature ferromagnetic and the colossal
 magnetoresistant phase under a magnetic field of 7 T). Robust
 signatures of the metal-insulator transition are identified in the
 momentum density for the paramagnetic phase above the Curie temperature.
 We interpret our results as providing direct evidence for the transition
 from the metallic-like to the admixed ionic-covalent bonding accompanying
the
 magnetic transition. The number of electrons involved in this phase
transition
 is estimated from the area enclosed by the Compton profile
 anisotropy differences. Our study demonstrates the sensitivity of
the Compton
 scattering technique for identifying the number and type of electrons
 involved in the metal-insulator transition.

 \end{abstract}

 \maketitle

 %
 %

 Recent developments in spin-based electronics or {\em spintronics} have
 revived the interest in manganites \cite{Katsnelson}. In particular, the
 bilayer manganites La$_{2-2x}$Sr$_{1+2x}$Mn$_2$O$_7$ (LSMO) show
 interesting colossal magnetoresistance (CMR) effects \cite{moritomo96,kubota} in a complex phase diagram
 with charge and orbital ordering \cite{wilkins}. As seen in Fig.~1 of Ref.~\onlinecite{kubota}, the
compounds in the doping
 range $x = 0.3 - 0.4$ show a metal-insulator
 transition (MIT) associated with the onset of long-range ferromagnetic (FM) order. The
 MIT occurs at a Curie temperature $T_c$ of about $125$ K (for $x = 0.35$)
 accompanied by colossal changes in the magnetoresistance \cite{mit1,mit2}.
 Above $T_c$, the phase
 diagram displays an insulating paramagnetic (PM) phase. The Mn-$3d$
 electronic states, which are responsible for these properties, split
 into $e_{g}$ and $t_{2g}$ contributions in the crystal field of the
 MnO$_6$ octahedron. The FM phase below $T_c$ and its metallic
 conductivity are usually explained on the basis of the double exchange
 (DE) mechanism \cite{de1}, where $e_g$ electrons hop between Mn sites
 through hybridization with the oxygen $2p$ orbitals and align the
 localized $t_{2g}$ spins by the exchange interaction.
 While the DE mechanism appears to capture the tendency towards
 ferromagnetism, it still remains unclear if oxygen orbitals should be explicitly
 included in the electronic degrees of freedom, or whether they can be integrated
 out as is often assumed in the standard models \cite{coey,dagotto}.

 Recent magnetic Compton scattering (MCS) studies \cite{cooper}
 of the manganite FM phase have shown how the occupation numbers of the
 $e_g$ states vary with doping \cite{koizumi01,mijnarends07} as well as
 temperature \cite{koizumi04,li04}. In addition, they have provided
 evidence for the coexistence of localized and itinerant $e_g$ magnetic
 electrons \cite{koizumi06,bba05} in the FM phase. MCS has also
 been used to study other spintronics materials such as magnetite
 Fe$_3$O$_4$ and its mysterious Verwey transition \cite{li07}.

 In this letter, we show that the anisotropy of high resolution Compton
 profiles (CP) displays a striking difference between the insulating
 PM and metallic FM case. This is important because this
 difference originates in the MnO planes which are the seat of
 the CMR properties. Similar effects have been observed in the
 metallic YBa$_2$Cu$_3$O$_7$ and insulating PrBa$_2$Cu$_3$O$_7$ systems
 \cite{shukla}. However, this is the first time that this effect has been
 observed on the same sample under the influence of external parameters
 such as temperature and magnetic field. We also provide a
 measure of the number of electrons involved in the CMR effect.

 Compton scattering, or inelastic scattering with very high momentum and
 energy transfer, is a probe of the ground state one-electron properties
 of the system \cite{cooper}. The measured one-dimensional quantity
 $J(p_z)$ is a projection of the three-dimensional electron momentum
 density $\rho(p_x,p_y,p_z)$ onto the $z$-axis, which lies along the
 scattering vector:
 \begin{equation}
 J(p_z) = \int \int \rho(p_x,p_y,p_z)
 dp_xdp_y
 \label{eq1}
 \end{equation}

 %
 %

 The sample used was a single crystal of
 La$_{2-2x}$Sr$_{1+2x}$Mn$_2$O$_7$ with $x = 0.35$, which was melt-grown
 in flowing oxygen gas in a floating zone optical furnace \cite{exp1}.
 According to the magnetic phase diagram determined by
 neutron-diffraction measurements, the present sample shows a
 ferromagnetic (FM) phase below $125$ K \cite{kubota}. High-resolution CP
measurements
 were carried out with a Cauchois-type x-ray
 spectrometer installed on the BL08W beam line at SPring-8, Japan
 \cite{exp1,exp2,exp3}. The energy of incident x-rays was $115.6$ keV and
 the scattering angle was $170\,^{\circ}$. The momentum resolution is
 estimated to be $0.15$ atomic units (a.u.).  CPs along the [100] and [110]
 axes were measured in three different conditions: the FM metallic state
 at low temperature ($20$ K), the CMR state in an external field of $7$ T,
 and the PM insulating state at $131$ K (above $T_c$).

 The electronic structure, momentum density, and the CPs along principal
 symmetry directions used for analyzing the present measurements
have been computed within an all-electron, fully charge
 and spin self-consistent semi-relativistic KKR framework \cite{bansil99}
 for LaSr$_2$Mn$_2$O$_7$ in the I4/mmm \cite{cryst} crystal
structure \cite{mijnarends07}. Our computed electronic
 structure (for $x=0.5$) is in good accord with
 other studies \cite{deboer99,huang00} The effects of doping $x$
 have been included within a rigid band model. All calculations agree on a
 nearly or wholly halfmetallic FM band structure with the
 Fermi level crossing Mn $d$ bands.

 %
 %
 %

 Since the LSMO electronic structure has a two-dimensional character, we
 shall focus on the calculated (001) 2D-projection of the momentum
 density given by
 \begin{equation}
 \rho^{2d}(p_y,p_z) = \int  \rho(p_x,p_y,p_z) dp_x~.
 \label{eq2}
 \end{equation}

\begin{figure}
 \begin{center}
 \includegraphics[width=\hsize]{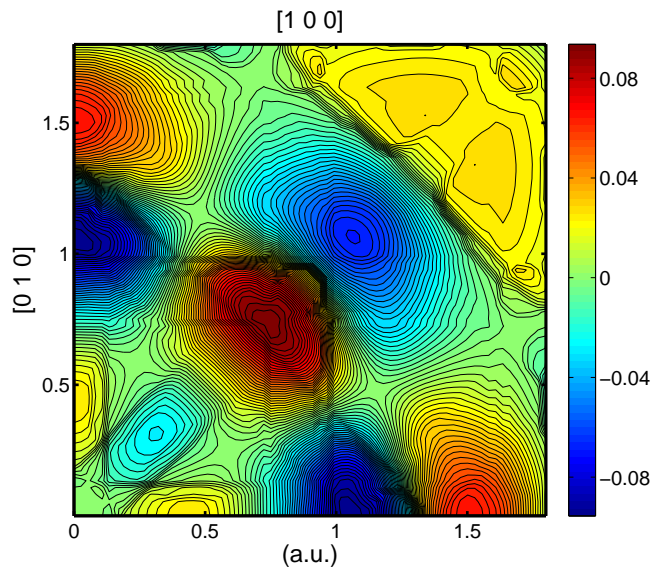}
 \end{center}
 \caption{(Color online)
 Anisotropy $A^{2d}_{C_{4v}}(p_y,p_z)$ of the  2D-projection
 of the theoretical electron momentum density onto the (001) plane, normalized to
the
 amplitude $\rho^{2d}(0,0)$. }
 \label{fig1}
 \end{figure}

 Apart from Fermi surface (FS) effects, the calculated distribution
 $\rho^{2d}(p_y,p_z)$ contains some anisotropic components produced by
 the wave functions of the $e_g$ and $t_{2g}$ electrons. We extract the
 $C_{4v}$ \cite{Tinkham} anisotropy from $\rho^{2d}(p_y,p_z)$ by taking the difference
 \begin{equation}
 A^{2d}_{C_{4v}}(p_y,p_z)=\rho^{2d}(p_y,p_z)-
 \rho^{2d}(\frac{p_y+p_z}{\sqrt{2}},\frac{p_y-p_z}{\sqrt{2}})~.
 \label{eq3}
 \end{equation}
 The subtraction in Eq.~\ref{eq3} acts as a projector on the $e_g$ and
 $t_{2g}$ subspace with the advantage of eliminating the large isotropic
 contribution of the core and some irrelevant valence electrons. The
 anisotropy $A^{2d}_{C_{4v}}(p_y,p_z)$ shown in Fig.~\ref{fig1} displays
 a strong peak near $(0.7, 0.7)$ a.u. which arises from wavefunctions of
 $t_{2g}$ symmetry. On the other hand, there are peaks on the [100] axes
 at a distance of about $1.5$ a.u. possessing $e_g$ character and smaller
 peaks in between $0.4$ and $0.5$ a.u. due to the oxygen $p$ states at
 the FS. These oxygen $p$ states hybridize with the $e_g$ states of Mn.
 The amplitude of the $t_{2g}$ and $e_g$ peaks is about $9$ \% of
 $\rho^{2d}(0,0)$ while the amplitude of the $p$ related peaks is less
 than $4$ \%. As $A^{2d}_{C_{4v}}(p_y,p_z)$ integrates to zero, it
 assumes positive and negative values.

 %
 %
 %

\begin{figure}
\begin{center}
\includegraphics[width=8cm,height=6cm]{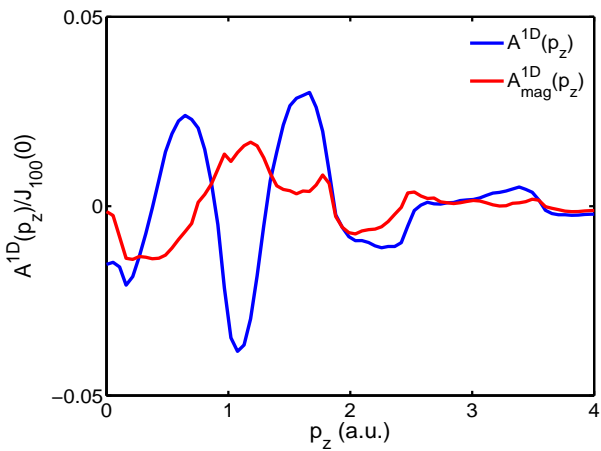}
\end{center}
\caption{(Color online) Theoretical anisotropies $A^{1D}(p_z)$ and
$A^{1D}_{mag}(p_z)$ \cite{mijnarends07} (without resolution
broadening).} \label{fig2}
\end{figure}

 Interestingly, Fig.~\ref{fig1} displays the same sharp FS
 features, which appear
 in the calculated 2D-projection of the spin momentum
 density onto the (001) plane \cite{li04}. In the present case,
 because of the definition of $A^{2d}_{C_{4v}}(p_y,p_z)$, we
 also have FS signatures rotated by $45\,^{\circ}$ with respect to the
 original FS: These features come from the subtraction of the rotated
 2-dimensional distribution in Eq. (3). The large nested portions of the
FS seen in
 Fig.~\ref{fig1} give maxima in the generalized charge susceptibility
 \cite{saniz} and explain the interplay of the conduction electrons and
 lattice vibrations which contribute to the MIT \cite{littlewood}.
 However, our aim here is not to investigate these FS properties in
 detail but rather to show how Compton spectroscopy may be used to study
 bonding effects across the MIT in the bilayer manganites. In order
 to analyze bonding effects we now discuss the CP anisotropy
 \cite{isaacs}, i. e. the difference
 between CPs measured
 in two crystallographic directions.
 In the present case, we consider
 \begin{equation}
 A^{1D}(p_z)=J_{100}(p_z) -J_{110}(p_z)~.
 \label{eq4}
 \end{equation}
 Recalling from Eq.~(\ref{eq1})
 that the CP involves a double integral, or
 equivalently, a 1D-projection of the momentum density,
 the shape of $A^{1D}(p_z)$ can be computed
 from the 1D-projection of
 the previous two-dimensional anisotropy
 $A^{2d}_{C_{4v}}(p_y,p_z)$
 as
 \begin{equation}
 A^{1D}(p_z)=\int A^{2d}_{C_{4v}}(p_y,p_z) dp_y~.
 \label{eq5}
 \end{equation}
 In Fig.~\ref{fig2}, the profile $A^{1D}(p_z)$
 shows the signatures of both the $t_{2g}$ peak (at
 about 0.7 a.u.) and the $e_g$ peak (at about 1.6 a.u.)
 discussed earlier and visible in Fig.~\ref{fig1}.
 Fig.~\ref{fig2} shows that at low momenta
 the anisotropy of the MCP
 $A^{1D}_{mag}(p_z)$ \cite{mijnarends07}
 (mostly of Mn-$d$ character)
 is markedly different from $A^{1D}(p_z)$.
 Clearly, the non-magnetic contribution
 due to the oxygen atoms is expected
 to give a more important impact at low momenta.

 %
 %

\begin{figure}
\begin{center}
\includegraphics[width=8cm,height=6cm]{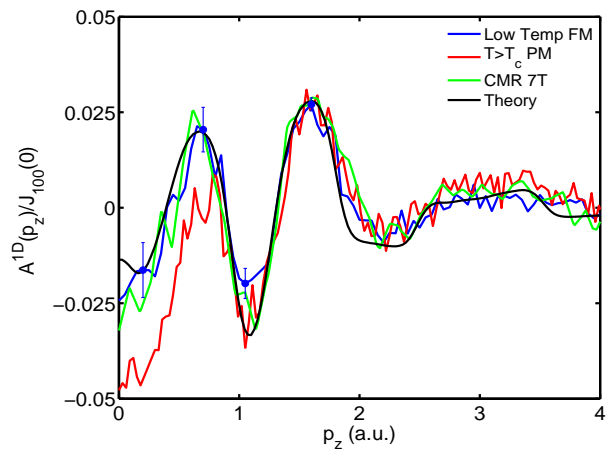}

\end{center}
\caption{(Color online) Comparison of the theoretical and
experimental $A^{1D}(p_z)$. In the legend, Low Temp FM is for the ferromagnetic
phase at $20$ K, $T> T_c$ is for the paramagnetic phase at $131$ K, and CMR $7$ T is for the colossal magnetoresistance phase in an external field of $7$ T at
$131$ K. Theory is convoluted with the experimental resolution.
Error bars shown for the FM phase are comparable for other
experimental CP anisotropies.} \label{fig3}
\end{figure}

 Next, we compare the calculated CP anisotropy with our measurements. In
 Fig.~\ref{fig3} all the {\em metallic} $A^{1D}(p_z)$ look remarkably
 alike, showing that the overall description of the electronic structure
 is satisfactory. In particular, the amplitude of $A^{1D}(p_z)$ in the
 resolution-broadened theory is the same as that in the experiment while
 in the case of the cuprates the theoretical $A^{1D}(p_z)$ had to be
 scaled down by a factor of $1.4$ to obtain agreement with experiment
 \cite{shukla}. The experimental curve for the {\em insulator} (paramagnetic phase), however,
 presents important changes in the low-momentum region indicating that
 the electrons with $p$ oxygen character at the FS structure near (0.5,
 0) a.u. in Fig.~\ref{fig1} are significantly affected by the MIT. Moreover, the anisotropy
 for the insulator has a more pronounced positive excursion for $p > 4$ a.u. This feature at high momenta is
rather small in amplitude but it makes the integrated anisotropy to
be zero.
 These
 changes in $A^{1D}(p_z)$ can be explained in terms of a transition from
 metallic to admixed ionic-covalent Mn-O bonding \cite{good}. According
 to the double-exchange mechanism \cite{de1}, if the oxygen ion separates
 ferromagnetically coupled Mn$^{3+}$ and Mn$^{4+}$ in a lattice of
 disordered Mn ions, the state Mn$^{3+}$-O-Mn$^{4+}$ is degenerate with
 the state Mn$^{4+}$-O-Mn$^{3+}$ so that the $d$ electrons develop a long
 range phase coherence by hopping between Mn sites \cite{de1}. Therefore,
 in the FM phase below $T_c$, the bonding is metallic while in the PM
 phase above $T_c$ the bonding becomes admixed ionic-covalent. The
 latter is described by a short-range phase coherence.

 %
 %

\begin{figure}[t]
\begin{center}
\includegraphics[width=8cm,height=6cm]{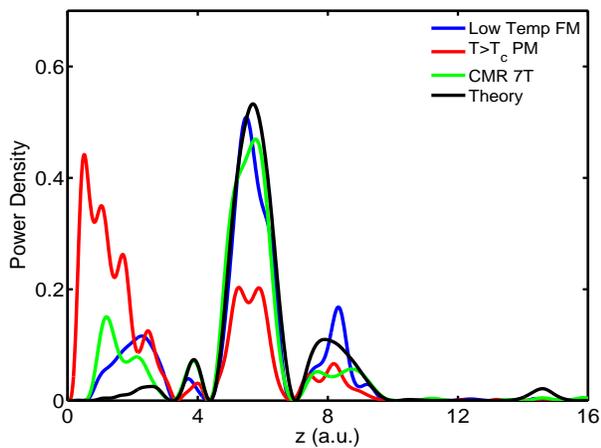}
\end{center}
\caption{(Color online) Comparison of the theoretical and
experimental $P_D(z)$. All the curves are normalized to a unit
area. The notation in the legend is as in Fig.~\ref{fig3}} \label{fig4}
\end{figure}

 The wavefunction localization effects across the MIT can also be
 monitored by studying the power density \cite{shukla}
 \begin{equation}
 P_D(z)= \left|\int dp_z A^{1D}(p_z) \exp(ip_{z}z)\right|^2.
 \label{eq7}
 \end{equation}
The power density of the anisotropy separates in real space the
different length scales contributing to the oscillations in the
anisotropy in momentum. Thus, the peaks in the power density
$P_D(z)$ indicate characteristic distances over which wave functions
are coherent in given crystallographic directions. Figure~\ref{fig4}
shows that the experimental data for the metallic phases and the
theory are in agreement and that the wave functions are of a
delocalized nature. However, the experiment for the insulator shows
a clear tendency to shift spectral weight towards short distances.
Thus, the localization trend for the insulator wave functions
becomes clear. To ease the comparisons we have normalized all the
power spectra to a unit area in Fig.~\ref{fig4} since only the
relative weights in the same spectrum are relevant to study the
localization\cite{shukla}.
 %
 %

 The number of electrons involved in this localization can
 be extracted from the area enclosed by the anisotropy difference
 \begin{equation}
 \Delta A^{1D}(p_z)= A^{1D}_{PM}(p_z)-A^{1D}(p_z),
 \label{eq8}
 \end{equation}
 where $A^{1D}(p_z)$ is a metallic anisotropy (FM or CMR)
 and $A^{1D}_{PM}(p_z)$ is an insulating anisotropy.
Therefore the quantity $n_e$ given by
 \begin{equation}
 n_e= \frac{1}{2}\int dp_z \left|\Delta A^{1D}(p_z) \right|~,
 \label{eq9}
 \end{equation}
measures the number of electrons whose wave functions change across
the MIT. This is due to the shift of plane wave components from low
to high momenta resulting in the wave function localization. We have
normalized $n_e$ to obtain the number of displaced electrons per Mn
atom. The calculated values of $n_e$ correspond to a displacement of
$0.71$ electron per Mn for the low-T FM-PM transition and of $0.67$
electron per Mn for the CMR-PM transition These values are
calculated from the anisotropies between $0$ and $9.6$ a.u. Their
error is $\pm 0.08$ based upon the statistical errors of the
profiles. Thus, the present MIT gives particularly strong and robust
signatures in momentum space.

 In conclusion, our study shows that high-resolution Compton
 scattering spectra display striking features of the MIT and provide
unique
 insights into the PM phase and the CMR effect in manganites.
Ferromagnetism and
 metallic conductivity in the CMR phase
 are explained in terms of $e_g$ long range
 phase coherence produced by the DE mechanism.
 The momentum density anisotropy reveals
 that both the Mn and the oxygen
 orbitals play a key role in the MIT. By measuring
 the number of displaced electrons $n_e$ across the MIT,
 the present method opens new opportunities for
 studying quantitatively the phase diagrams of
 complex materials.

 We acknowledge discussions with Robert Markiewicz. This work was
 supported by a Grant-In-Aid for Science and Culture, Japan and by
 contract No. DE-FG02-07ER46352 from the Division of Materials
 Science and Engineering, Office of Science, U.S. Department of
 Energy. It benefited from the allocation of supercomputer time at NERSC,
the
 Northeastern University's Advanced Scientific Computation Center (ASCC),
 and the Stichting Nationale Computerfaciliteiten (National Computing
 Facilities Foundation, NCF). The Compton scattering experiments were
 performed with the approval of the Japan Synchrotron Radiation Research
 Institute (JASRI) (Proposal No. 2002B2008, 2003A3008, 2003B4008, and
 2004A5008-LD3-np).

 \begin{noindent}

  \end{noindent}

  \end{document}